\documentclass{elsart}
\usepackage[dvips]{graphicx}
\usepackage[french]{babel}
\usepackage[latin1]{inputenc}
\begin{document}
\begin{frontmatter}
\title{Crise d'ébullition : inhibition du détachement de la bulle de vapeur par la force de recul}
\author[SBT]{Vadim S. Nikolayev\thanksref{email}}
\author[SBT]{Daniel Beysens}\and \author[CNRS]{Yves Garrabos}
%% Adresse
\address[SBT]{ESEME, Service des Basses Températures, CEA Grenoble, France\thanksref{mail}}
\thanks[email]{Email: vnikolayev@cea.fr}
\thanks[mail]{Adresse postale: CEA-ESEME, PMMH-ESPCI, 10, rue Vauquelin
75231 Paris Cedex 05, France }
\address[CNRS]{CNRS-ESEME,Institut de Chimie de la Mati\`ere Condens\'{e}e de Bordeaux,\\
87, Av. du Dr. Schweitzer, 33608 Pessac Cedex, France }

\begin{abstract}La crise d'ébullition est une transition entre deux régimes d'ébullition :
ébullition nucléée (la bulle se forme sur la surface chauffante) et ébullition en film (la surface chauffante
est couverte par un film continu de vapeur séparant la surface chauffante du liquide). Dans cette
communication, nous présentons un modèle physique de la crise d'ébullition basé sur le concept de recul de
vapeur. Nos simulations numériques de croissance thermiquement contrôlée montrent comment une bulle attachée
à la surface chauffante commence soudainement à s'y étaler, formant le germe d'un film de vapeur. La force de
recul de vapeur ne provoque pas seulement l'étalement de la bulle, elle crée également une force
additionnelle d'adhérence qui empêche le départ de la bulle de la surface chauffante lors de sa croissance.

Près du point critique liquide-vapeur, la croissance de la bulle est très lente. Si, de plus, des conditions
de microgravité sont remplies, la bulle garde la forme convexe et il est possible d'observer
expérimentalement une augmentation de l'angle apparent de contact ainsi que la croissance de la tache sèche.
Ces observations confirment l'explication proposée.

\vspace*{\baselineskip} \noindent\textbf{Abstract}

\vspace*{\baselineskip} \noindent Boiling crisis is a transition between nucleate and film boiling. In this
communication we present a physical model of the boiling crisis based on the vapor recoil effect. Our
numerical simulations of the thermally controlled bubble growth at high heat fluxes show how the bubble
begins to spread over the heater thus forming a germ for the vapor film. The vapor recoil force not only
causes the vapor spreading, it also creates a strong adhesion to the heater that prevents the bubble
departure, thus favoring the further bubble spreading.

Near the liquid-gas critical point, the bubble growth is very slow and allows the kinetics of the bubble
spreading to be observed. Since the surface tension is very small in this regime, only microgravity
conditions can preserve a convex bubble shape. Under such conditions, we observed an increase of the apparent
contact angle and spreading of the dry spot under the bubble, thus confirming our model of the boiling
crisis.

\vspace*{\baselineskip} \noindent\textit{Mots-clefs : } crise d'ébullition ; flux critique ; point critique ;
microgravité
\end{abstract}
\end{frontmatter}

\section{Introduction}

Quand le flux de chaleur provenant d'une paroi chauffante dépasse une valeur critique pendant l'ébullition
(le flux critique, le CHF en anglais), la vapeur forme brutalement un film sur la paroi chauffante, qui
l'isole thermiquement du liquide. En d'autres termes, la paroi chauffante se dessèche. Le transfert thermique
est inhibé et la température de la paroi croît rapidement.  Ce phénomène s'appelle la crise d'ébullition par
caléfaction \cite{Tong}. L'évaluation correcte du CHF exige une compréhension claire du phénomène physique
qui déclenche la crise. De nombreux modèles sont proposés, cf. \cite{Buyevich} pour une revue. Pourtant,
chacun d'eux ne s'applique qu'à des régimes et des configurations d'expériences très particuliers. Pour des
raisons d'importance industrielle, les expériences d'ébullition les plus communes sont effectuées sous
pesanteur terrestre à des pressions basses par rapport à la pression critique du fluide donné. La valeur du
CHF est alors grande, ce qui rend l'ébullition violente et les observations difficiles. Même les dispositifs
les plus avancées \cite{Theo} ne peuvent aider à valider un modèle avec certitude. Cependant, de nombreuses
observations montrent que la crise commence par la croissance d'une tache sèche sous les bulles de vapeur
attachées à la paroi.

En diminuant la force d'Archimède, la microgravité permet d'améliorer la qualité des observations grâce au
temps de résidence plus long de la bulle sur la paroi chauffante \cite{Straub}. Cependant, la croissance
reste aussi rapide que sur Terre. Sous de grandes pressions, la croissance est plus lente, l'influence des
flux hydrodynamiques sur cette croissance est moins importante, ce qui permet d'identifier les mécanismes
thermiques, notamment l'influence de la force de recul de vapeur sur la croissance de la tache sèche. Du
point de vue de la modélisation, cela permet une grande simplification en négligeant (en première
approximation) le couplage thermique-hydrodynamique et en supposant une forme quasi-statique de la bulle.
Dans la section \ref{sec-f} ci-dessous, nous développons l'approche  \cite{EuLet99} qui montre que la force
de recul peut être à l'origine de la crise d'ébullition.

Si les pressions sont si grandes que le fluide se trouve près de son point critique, la croissance de la
bulle devient exceptionnellement lente et permet des observations très claires de l'assèchement de la paroi
\cite{gaswets}, cf. section \ref{sec-mg} ci-dessous.

\section{Force de recul de vapeur et croissance de la tache sèche sous une bulle}\label{sec-f}

La force de recul de vapeur vient de l'impulsion mécanique non compensée des molécules qui quittent
l'interface liquide-gaz lors de l'évaporation. Sa valeur par unité de surface (= pression) s'écrit
\cite{Palm} $P_r=\eta^2(\rho_V^{-1}-\rho_L^{-1})$ où $\eta$ est le taux massique d'évaporation, c.-à-d. la
masse de fluide évaporée par unité de surface pendant l'unité du temps, et $\rho_L$($\rho_V$) signifie la
densité du liquide (vapeur).  La force de recul est normale à l'interface et est toujours dirigée vers le
liquide. En négligeant la conductivité thermique dans la vapeur par rapport à celle dans le liquide, on écrit
$q_L=H\eta$ où $H$ est la chaleur latente de vaporisation et $q_L$ et le flux thermique qui arrive à
l'interface du coté liquide.

\begin{figure}[htb] \centering
\includegraphics[width=7cm]{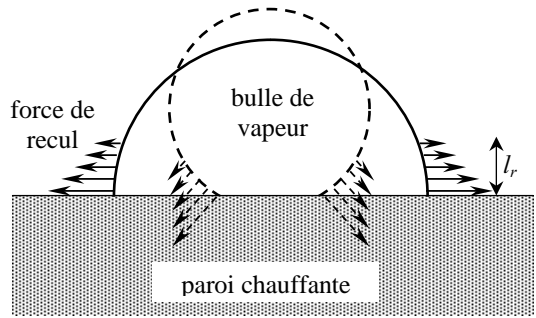}
\caption{Etalement d'une bulle de vapeur sous l'action de la force de recul de la vapeur.} \label{brec}
\end{figure}

Considérons maintenant une bulle de vapeur attachée à la surface de la paroi chauffante (Fig.~\ref{brec}).
Lors de l'ébullition, le liquide est surchauffé dans une couche pariétale d'épaisseur $l_r$. Cependant, la
température de l'interface vapeur-liquide est constante (c'est en fait la température de saturation pour la
pression donnée du système). Cela signifie que le flux $q_L$ reste élevé dans une ``ceinture" sur la surface
au pied de la bulle. De fait, la majeure partie de l'évaporation dans la bulle de vapeur est produite dans
cette ceinture, dont l'épaisseur est habituellement beaucoup plus petite que le rayon de la bulle
\cite{Carey}.  Le recul de la vapeur près de la ligne de contact est alors beaucoup plus grand que sur
l'autre partie de la surface de la bulle. En conséquence, sa surface se déforme comme si la ligne triple de
contact liquide-vapeur-solide était tirée de coté, comme montré dans la Fig.~\ref{brec}. Ceci signifie que,
sous l'action du recul de la vapeur, la tache sèche sous la bulle de vapeur s'étale et peut couvrir la
surface de la paroi chauffante.

Dans ce qui suit, nous montrons par un exemple simple que l'influence du recul de vapeur peut être
interprétée en terme de changement d'angle apparent de contact.

\subsection{Recul de vapeur et angle de contact apparent: une approche simplifiée}

\begin{figure}[htb] \centering
\includegraphics[width=8cm]{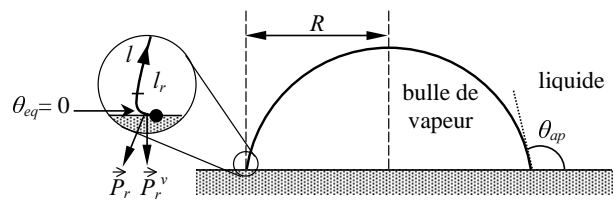}
\caption{Les angles de contact apparent $\theta_{ap}$ et réel $\theta_{eq}$.  Le point noir indique la ligne
de contact, c.-à-d. zéro pour la coordonnée curviligne $l$.} \label{appangle}
\end{figure}

Considérons une coordonnée curviligne $l$ mesurée le long du contour de la bulle, comme dans la
Fig.~\ref{appangle}. Deux forces définissent la forme de la bulle dans l'approximation quasi-statique : la
force de recul et la tension superficielle $\sigma$,
\begin{equation}
K\sigma=\lambda+P_r(l),\label{surf}
\end{equation}
où $K$ est la courbure locale de la bulle et $\lambda$ est la différence de pression entre l'intérieur et
l'extérieur de la bulle, indépendante de $l$. Cette équation est difficile à résoudre analytiquement et le
calcul numérique \cite{EuLet99,IJHMT01} (onéreux au niveau du temps) s'impose. Cependant, l'approche
simplifiée ci-dessous permet d'estimer et de mieux comprendre l'effet de la force du recul sur la forme de la
bulle.

Par angle de contact apparent $\theta_{ap}$, nous sous-entendons celui mesuré à la distance $l_r$ (=
épaisseur de la couche limite) de la ligne triple comme marqué dans la Fig.~\ref{appangle}. Cette définition
est valable quand $l_r\ll R$, le rayon de la bulle. Dans ce cas-là, la fonction $P_r(l)$ peut être approximée
par la fonction $\delta$ de Dirac:
\begin{equation}\label{delta}
P_r(l)=2\sigma_r\delta(l).
\end{equation}
La fonction $\delta(l)$ est définie de telle manière, que $\delta(l)=0$ si $l\neq 0$, $\delta(0)=\infty$  et
$\int_{-\infty}^\infty\delta(l)\textrm{d}l=1$. D'après (\ref{surf}), ceci signifie que le recul de vapeur est
localisé à la ligne de contact, et la tension superficielle donne au reste de la surface de la bulle la forme
d'une calotte sphérique. La supposition (\ref{delta}) se justifie par le calcul plus rigoureux \cite{EuLet99}
qui montre que $P_r(l)$ diverge quand $l\rightarrow 0$, $\int_0^\infty P_r(l)\textrm{d}l$ restant fini.

\begin{figure}[htb]
\centering
\includegraphics[width=5cm]{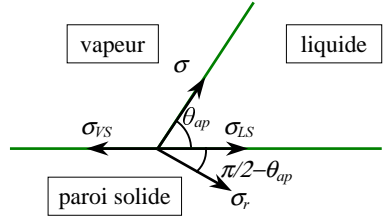}
\caption{Equilibre des forces qui agissent sur la ligne triple de contact, où $\sigma_{vs}$ et $\sigma_{ls}$
sont les tensions superficielles pour les interfaces respectivement vapeur-solide et liquide-solide.}
\label{Young}
\end{figure}

L'amplitude $\sigma_r$ du recul de vapeur a la dimension d'une tension superficielle et doit être
incluse\footnote{En se basant sur l'approche variationelle \cite{EuLet99}, on peut montrer cela
rigoureusement.} dans le bilan de (quasi-)équilibre des tensions agissant sur la ligne de contact montrées
dans Fig.~\ref{Young}.
La ligne de contact est stationnaire quand la composante horizontale de la somme des vecteurs de toutes les
forces est égal à zéro, ce qui se résume en
\begin{equation}\label{theta-ap}
  \cos\theta_{ap}=\cos\theta_{eq}-N_r\sin\theta_{ap},
\end{equation}
où l'angle de contact d'équilibre est donné par l'expression classique
$\cos\theta_{eq}=(\sigma_{vs}-\sigma_{ls})/\sigma$.  Le "coefficient de recul"
\begin{equation}\label{Nr}
  N_r={\sigma_r\over\sigma}={1\over\sigma}\int\limits_0^{l_r}\,P_r(l)\,{\rm d}l,
\end{equation}
caractérise la force de recul de la vapeur relativement à la tension superficielle.

\begin{figure}[htb]
\centering
\includegraphics[width=8cm]{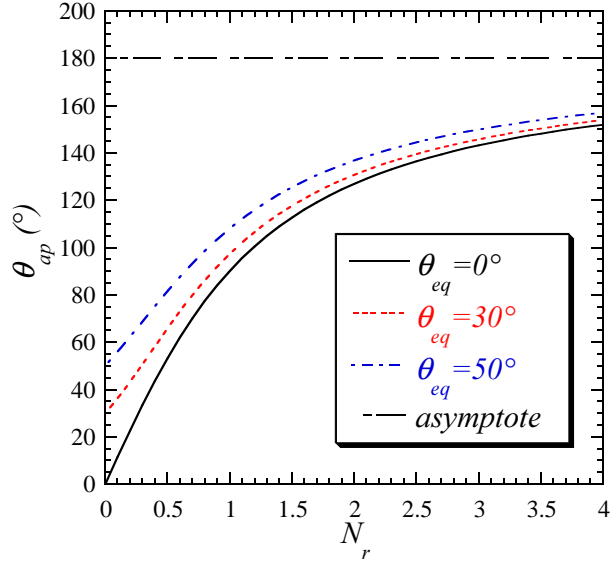}
\caption{Angle apparent de contact par rapport en coefficient de recul $N_r$ pour différentes valeurs de
l'angle de contact d'équilibre.} \label{theta-nr}
\end{figure}

La dépendance de $\theta_{ap}$ par rapport à $N_r$, calculée à partir de (\ref{theta-ap}), est présentée sur
la Fig.~\ref{theta-nr}.
On peut voir que la force de recul augmente l'angle de contact apparent.  Quand la puissance de la paroi
chauffante reste constante, le flux de la chaleur $q_L$ augmente avec le temps en augmentant $N_r$. Par
conséquent, l'angle de contact apparent croît dans le temps. Comme la bulle est une calotte sphérique (ce qui
est définie par l'eq.~(\ref{surf}) avec $P_r(l)=0, l\ne 0$), l'augmentation de l'angle de contact se traduit
par l'étalement de la tache sèche, son rayon $R_{dry}$ étant lié au rayon de la bulle $R$ (cf.
Fig.~\ref{appangle}) par l'expression \begin{equation}\label{rd}R_{dry}=R\sin\theta_{ap}
\end{equation} qui provient de la
géométrie de la calotte sphérique.

\begin{figure}[htb]
\centering
\includegraphics[width=8cm]{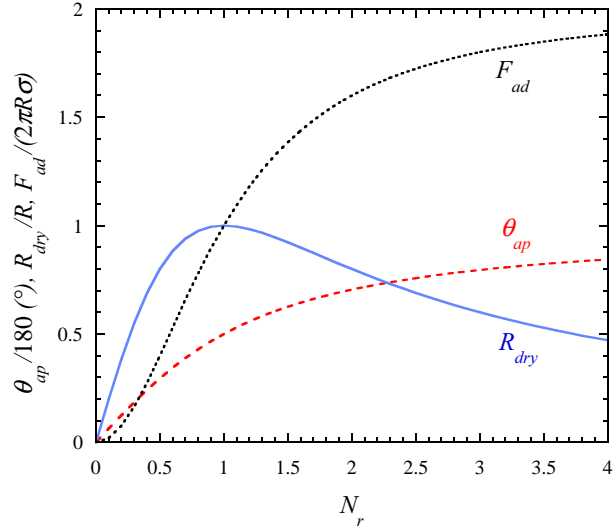}
\caption{Dépendance du rayon de la tache sèche et de la force d'adhésion du coefficient du recul $N_r$
calculés pour $\theta_{eq}=0$. $\theta_{ap}$ est aussi représenté pour comparaison.} \label{Simple_Nr}
\end{figure}

La force d'adhésion de la bulle à la paroi $F_{ad}$ s'avère très importante pour estimer la valeur du CHF
(voir ci-dessous). Elle peut être obtenue à partir du bilan vertical des tensions dans la Fig.~\ref{Young}.
La force $F_{ad}$ est le produit de la longueur( $=2\pi R_{dry}$) de la ligne triple  et de la tension
non-compensée :
\begin{equation}\label{fads}
F_{ad}=2\pi R\sigma(\sin\theta_{ap}-N_r\cos\theta_{ap})\sin\theta_{ap}.
\end{equation}
Les dépendances de $R_{dry}$ et $F_{ad}$ de $N_r$ sont montrées dans la Fig.~\ref{Simple_Nr}.
Le comportement de la force d'adhésion est correcte: elle augmente avec $N_r$. En contrepartie, la
décroissance de $R_{dry}$ donnée par (\ref{rd}) à partir d'une certaine valeur de $N_r$ est paradoxale. Cette
décroissance apparaît quand $\theta_{ap}$ dépasse 90°. La composante verticale $\vec{P}_r^v$ de $\vec{P}_r$
change alors de direction. Cette incohérence est due au fait que la forme exacte de la bulle dans la
proximité de la ligne triple (cf. l'inséré dans Fig.~\ref{appangle}) n'est pas prise en compte. En réalité,
la surface est fortement courbée et $\vec{P}_r^v$ est dirigée vers le bas. Un autre défaut de la théorie
simplifiée réside dans l'impossibilité de calculer le temps de départ (puisque avant le départ, la forme de
la bulle dévie fortement de celle d'une calotte sphérique) C'est pourquoi il est nécessaire d'abandonner la
notion d'angle apparent qui est fondée sur l'expression (\ref{delta}) pour la force de recul. Cette dernière
doit être obtenue plus rigoureusement pour calculer la forme de la bulle, ce qui sous-entend une approche
numérique ou des expériences.

Cependant, l'approche de l'angle apparent ci-dessus illustre bien l'idée principale de l'étale\-ment sous
l'action du recul. Elle permet, en principe, de calculer le champ thermique autour de la bulle en tenant
compte de l'effet de recul sans traiter l'eq. (\ref{surf}) pour déterminer la forme de la bulle. On peut en
effet effectuer le calcul thermique pour une calotte sphérique et ajuster son angle de contact en utilisant
l'eq. (\ref{theta-ap}).

\subsection{Résultats d'une approche numérique}

\begin{figure}[htb] \centering
\includegraphics[width=8cm]{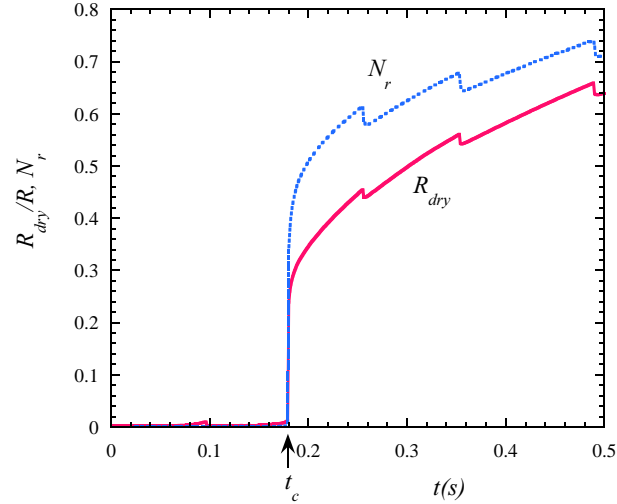}
\caption{Evolutions temporelles du rayon $R_{dry}$ de la tache sèche et du coefficient du recul $N_r$ issus
du calcul réalisé d'après l'article \cite{IJHMT01} pour $q_0=0.1$~MW/m$^2$. } \label{rdry}
\end{figure}

Les simulations numériques \cite{IJHMT01}, qui sont basées sur l'eq.~(\ref{surf}) avec $P_r(l)$ calculée à
partir de la distribution de température dans le liquide et la paroi chauffante, mettent en évidence la
croissance de la tache sèche. La précision des calculs est améliorée par rapport à l'article cité en
utilisant un nouvel algorithme \cite{BEM24}. La tache sèche reste petite jusqu'au temps de transition $t_c$
(cf. Fig.~\ref{rdry}) où elle s'agrandit brusquement. Nous associons cette transition avec la crise
d'ébullition. La croissance de la tache sèche juste avant la crise est confirmée par nombre de travaux
expérimentaux, voir par exemple \cite{Theo}. Etant très petite jusqu'à l'instant $t_c$, la force de recul
caractérisée par le paramètre $N_r$ devient alors de l'ordre de l'unité (cf. Fig.~\ref{rdry}) ce qui se
compare bien avec l'estimation pour le CHF donnée dans l'article \cite{EuLet99}. Abordons maintenant la
question de la valeur du CHF.

\subsection{Comment estimer le CHF: inhibition du détachement de la bulle par la force de recul}

\begin{figure}[htb]
\centering
\includegraphics[width=8cm]{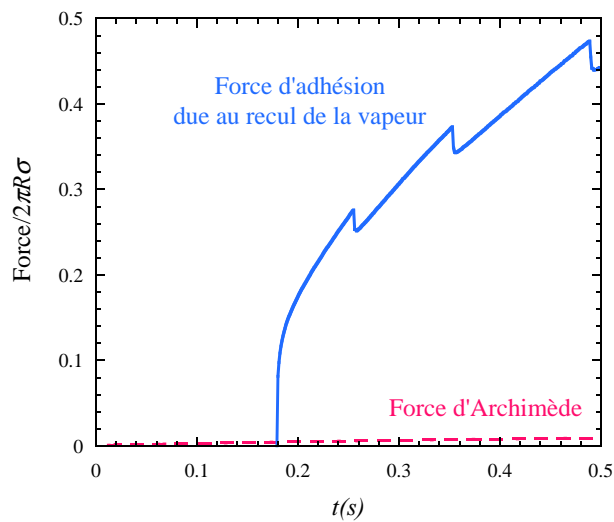}
\caption{Evolutions temporelles de la force d'Archimède et de la force d'adhésion $F_{ad}^r$, issues du
calcul réalisé d'après l'article \cite{IJHMT01} pour un flux $q_0=0.1$~MW/m$^2$. Les forces sont exprimées
dans les unités $2\pi R\sigma$ pour pouvoir comparer avec les résultats de l'approche simplifiée
(Fig.~\ref{Simple_Nr}).} \label{dep}
\end{figure}

Comme les "forces de départ" (force d'Archimède, poussée du flux hydrodynamique,...) qui tendent à arracher
la bulle de la paroi sont absentes de l'eq.~(\ref{surf}), la bulle reste toujours accrochée sur la paroi dans
la simulation de type \cite{IJHMT01}. Dans la situation réelle, la bulle quitte la paroi au moment $t_{dep}$.
Si $t_{dep}$ est plus petit que le temps $t_c$ issu de la simulation \cite{IJHMT01}, la bulle quitte la paroi
avant que la tache sèche puisse s'étaler et la crise d'ébullition ne se produit pas. L'analyse de départ de
la bulle s'avère alors cruciale pour trouver la valeur du CHF.

Le temps $t_c$ est une fonction décroissante du flux thermique $q_0$ de paroi \cite{IJHMT01}. Cependant,
$t_{dep}$ dépends aussi de $q_0$. Pour trouver la fonction $t_{dep}(q_0)$, on doit analyser la force
d'adhésion $F_{ad}$ d'une bulle à la paroi. Cette force se divise en deux parties
$F_{ad}=F_{ad}^\sigma+F_{ad}^r$. La première partie s'écrit $F_{ad}^\sigma=2\pi
R_{dry}\sigma\sin\theta_{eq}$. $F_{ad}^\sigma$ est très importante au début de la croissance, lorsque la
ligne triple est attachée au défaut de la paroi qui a servi de centre de nucléation (et qui donne à
$\theta_{eq}$ une grande valeur), elle devient négligeable par la suite. Comme on ne peut pas connaître la
valeur locale de $\theta_{eq}$, $F_{ad}^\sigma$ est difficile à estimer. On suppose ensuite que
$\theta_{eq}=0$ sur le reste de la paroi où la ligne triple se déplace. La deuxième partie
\begin{equation}\label{fad}
F_{ad}^r=\int P_r^v\textrm{d}A
\end{equation} est due à la force de recul dont la composante verticale $P_r^v$ (dirigée
vers la paroi) est intégrée sur toute la surface $A$ de la bulle. Bien que le calcul \cite{IJHMT01} est
bidimensionnel, on peut raisonnablement supposer que la valeur de la $P_r$ obtenue s'applique aussi pour le
cas 3D axi-symétrique (comme si la bulle avait une symétrie axiale en 3D).

Le calcul montre que $F_{ad}^r$ reste très petite au début de l'évolution (ce qui correspond au cas
$t_{dep}<t_c$), par rapport aux forces de départ, notamment la force d'Archimède. Si la bulle reste attachée
à la paroi pendant ce régime, c'est grâce à la force $F_{ad}^\sigma$. A partir de l'instant $t=t_c$, la
croissance de $F_{ad}^r$ s'accélère brutalement et c'est elle qui domine. Autrement dit, quand la tache sèche
commence à croître, la bulle reste attachée à la paroi en coalescant avec les bulles voisines. C'est ainsi
que la crise d'ébullition devrait se dérouler.

On voit (Fig.~\ref{dep}) que la croissance de la bulle est une superposition d'un processus monotone et
d'oscillations d'amplitudes croissantes. Ces oscillations apparaissent probablement à cause de l'instabilité
de recul \cite{Palm,Stein}. Une de ces oscillations initie la croissance brutale de la tache sèche.

Le comparaison des résultats numériques avec ceux du modèle simplifié montre que la force d'adhésion est
surestimée par ce dernier d'à peu près 50\%. Ce fait confirme la nécessite de poursuivre des études
numériques qui permettront d'obtenir la valeur du CHF compte tenu des forces de départ. Le CHF apparaît ainsi
comme la valeur du flux thermique de transition entre deux régimes qui ne peuvent pas coexister: le régime de
départ et le régime d'étalement de la bulle .

\section{Évidence expérimentale de l'étalement d'une bulle en microgravité}\label{sec-mg}

Les expériences \cite{gaswets} confirment que notre modèle d'étalement d'une bulle est valide dans la région
critique, c.-à-d. pour des pression et température proches de la pression et de la température critiques pour
le fluide donné.  Le point critique présente beaucoup de propriétés singulières.  En particulier, le
coefficient de diffusion thermique s'évanouit, ce qui ralentit la croissance des bulles et permet d'observer
les détails de la croissance sans une agitation gênante du liquide provoquée par des mouvements rapides. Par
exemple, la croissance d'une seule bulle a pu être observée \cite{gaswets} pendant environ quinze minutes.

\begin{figure}[hbt]
\centering
\includegraphics[width=8cm]{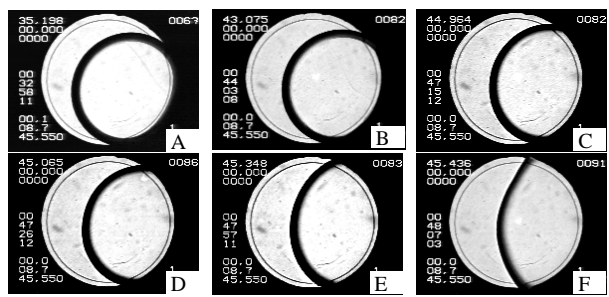}
\caption{Evolution d'une bulle de vapeur dans une cellule cylindrique remplie de SF$_6$ proche de son point
critique. Les images ont été obtenues pendant un chauffage continu de la cellule.  La dernière image (F)
correspond à la température juste en dessous de la température critique.  L'augmentation de l'angle de
contact apparent est évidente.} \label{img}
\end{figure}

Puisque la longueur capillaire $[\sigma/(\rho_L-\rho_V) g]^{1/2}$ (où $g$ est la gravité terrestre) disparaît
également au point critique et les bulles sont écrasées contres des parois. Les bulles de vapeur de forme
habituelle convexe ne sont donc pas observables en pesanteur terrestre. Pour cette raison, notre expérience a
été exécutée à bord de la station spatiale MIR. La valeur du CHF disparaît au point critique \cite{Tong}, par
conséquent, on s'attend à ce qu'un taux de chauffage très lent produise l'étalement de la bulle.  En raison
de cette évolution lente, les effets hydrodynamiques sont très faibles et la forme de la bulle n'est pas
distordue. Une cellule expérimentale de forme cylindrique a été remplie par du SF$_6$ à densité presque
critique.  On a observé l'évolution de la bulle à travers les bases transparentes du cylindre (fenêtres).  La
Fig.~\ref{img} montre comment la bulle, initialement circulaire ($\theta_{eq}=0$) s'étale sur la paroi
chauffante cylindrique.
L'angle de contact apparent augmente clairement avec le temps.  Bien que le volume de la bulle soit constant,
la masse de vapeur augmente suite à la densité croissante.  À la différence du régime de basse pression, la
densité dépend fortement de la température dans la région proche du point critique.

\section{Conclusions}

Nous proposons une explication physique pour la crise d'ébullition : l'étalement de la tache sèche sous une
bulle de vapeur est provoquée par la force de recul de la vapeur. Une fois l'étalement commencé, la bulle
reste attachée à la paroi par le même effet de recul et l'étalement (accompagné des coalescences possibles
avec des bulles voisines) peut se prolonger jusqu'à la formation du film continu de vapeur. L'approche
théorique est soutenue par la simulation numérique de la croissance de la bulle sous un grand flux de
chaleur, ainsi que par des données expérimentales sur la croissance de la bulle dans des conditions proches
du point critique. Des simulations numériques basées sur cette approche devraient permettre de déterminer le
CHF comme étant le flux thermique de transition entre deux régimes: le régime de départ et le régime
d'étalement de la bulle.

\end{document}